\documentclass[article]{revtex4}

\usepackage[dvips]{graphicx}

\begin{document}

\title{Dynamic density functional study of driven colloidal particles: the effect of the system dimension  }

\author{F. Penna and P. Tarazona }
\affiliation{%
Departamento de F{\'{\i}}sica Te\'orica de la Materia
Condensada,
Universidad Aut\'onoma de Madrid, E-28049 Madrid, Spain}
\date{\today}

\maketitle
The dynamical properties of classical fluids at  pico-liter scale 
 attract experimentally and theoretically  much attention in the soft-matter and biophysics communities, due to the appearance of the microfluidics ref.\cite{mic}, also called 'lab-on-a-chip' ref.\cite{labchip}, as one of the most relevance $21$th century technologies. Theses devices  try to put a full chemical laboratory onto your hand thought a sophisticated dispositive that includes  particles pump ref.\cite{pp}  and gates, chemical reaction  ref.\cite{Yeomans}, etc.. In this work we focus our attention on one of the key factors for the design and construction of  this type of devices: {\it the system dimension}. We consider a generic system formed by a dilute solution of  colloidal particles dragged ( at a constant rate c in the ${\bf \hat{z}}$ direction) by a moving potential barrier, modeled by a time dependent  external potential ( $V_{ext}({\bf r},t)=V_{ext}({\bf r}-ct{\bf \hat{z}})$) acting on the colloidal particles but with no effect on the solvent.
We base our results  on a  new technique named  Dynamic density functional (DDF) theory  ref.\cite{ddf,ddf2}   which  is a generalization of Density functional  (DF)  theory to out of equilibrium, for systems with relaxative molecule dynamics (those in which hydrodynamic modes are not relevant),  under the  hypothesis that the   dynamic correlations can be approximated by those  in a  system with the same density distribution at   equilibrium; the DDF is a promising way of describe classical fluids dynamics at a microscopic level including our knowledge of the very well developed DF theory for the equilibrium.\\

The central deterministic  DDF equation is,
\begin{eqnarray}
\frac{\partial \rho ({\bf r},t)}{\partial t}= \Gamma_o
\ \nabla \left[ \rho({\bf r},t) \ \nabla
\left( \frac{\delta{ \cal F}[\rho]}
{\delta \rho({\bf r},t)} +V_{{\rm ext}}({\bf r},t) \right) \right];
\label{DDF}
\end{eqnarray}
where $ \cal F$ is the Helmholtz free energy in DF theory, usually split onto the ideal gas contribution and the excess due to the interactions ($\Delta \cal F$), and $\Gamma_o$ is the mobility of an isolated colloidal particle in the solvent. \\
In the case of the shifting external potential we are dealing with, $V_{{\rm ext}}({\bf r}-ct{\bf \hat{z}})\equiv V_{{\rm ext}}( {\bf r'}) $  (where  ${\bf r'}$ is the coordinate in the reference framework of the external potential),  the most relevant feature comes from the stationary state (achieved after a transient period  from any initial state ) in which $\rho({\bf r},t)= \rho({\bf r}-ct{\bf \hat{z}})\equiv \rho( {\bf r'} )$ i.e. when the density distribution is shifted at the same rate c as the potential barrier,   so that,  eq.(\ref {DDF}) reduces to,

\begin{eqnarray}
\ \nabla \left[ \rho({\bf r'}) \ \nabla
\left( \frac{\delta{ \cal F}[{\rho}]}
{\delta \rho({\bf r'})} +V_{{\rm ext}}({\bf r'})+\frac{cz'}{ \Gamma_0
} \right) \right]=0.
\label{SDDF}
\end{eqnarray}
For non interacting particles, such that $\Delta {\cal F} =0 $,  eq.(\ref {SDDF}) becomes a linear Fokker-Planck
equation $\nabla^{2} \rho +
 \nabla(\rho \cdot \nabla\beta V_{k})=0 $, with a 
$"$kinetic potential$"$  given by  
$\beta V_{k}({\bf r'}) \, 
=\beta V_{{\rm ext}}({\bf r'})+\bar{c}z\,'$,
as a function of  the reduced shifting rate $\bar{c}\equiv \beta c/\Gamma_o$,
with inverse length units and $\beta=\frac{1}{k_{b}T}$. In this case, the effective dimension behaviour   is completely determined  by the difference between the system dimension (container) and the dimensions over which the external potential is homogeneous, so that an effective $1D$ system   got thought a   $2D$ plane barrier dragging particles in a $3D$ container, is equivalent to a point-like barrier dragging particles in  a narrow channel ref.\cite{PRLcha}; similarly, an effective  $2D$   system  may be observed  for a cylindrical barrier in a $3D$ container, or with  a point-like barrier in a  $2D$ container. For $\Delta {\cal F} \neq 0$ (interacting particles) ref.\cite{1d,joe2,prlfrus}  this is not longer true, because the excess free energy depends explicitly on the dimension of the  colloidal particles container. 

Since the system dynamic behaviour is determined mainly through the particles interaction and the effective system dimension, and  we are interested here in the characteristics induce by this last factor,  we  consider in this work only non interacting particles  which would represent diluted solutions of colloidal particles. We  set the notation before we start solving eq.(\ref {SDDF}) for each dimension. In the following, we refer to the {\it front} ({\it wake}) region as that away from the external potential, well beyond (behind) the z' positions with $V_{ext}\neq0$; if we assume that this external potential barrier is located around the  origin of the shifting reference framework, and restricted to values  $|z'|\approx \sigma$ with $\sigma$ the molecular diameter, then the points with $z'\gg\sigma$ are in the   {\it front}  while those with $z'\ll-\sigma$ are in the  {\it wake}. In both regions  $V_{ext}=0$, eq.(\ref{SDDF}) reduces to $\nabla^{2} \rho +\bar{c} \frac{\partial  \rho}{\partial z'} = 0 $, and the analytic form of $\rho({\bf r'})$ may be obtained as a series with the appropriate asymptotic limit in each  dimension.
Notice that only for effective $1D$ systems, the  {\it front}  and the  {\it wake} are fully separated by the external potential. In higher dimensions, they  are connected through those regions with $|z'|\leq \sigma $ but $|{\bf r'}|\gg \sigma $, for which  $V_{ext}=0$.
Our result are presented as follow: first we solve analytically eq.(\ref {SDDF}) (where $V_{ext}=0$),  them we find numerically the full solution, and finally we compare both.\\ 

For a shifting repulsive barrier in effective $1D$ systems ref.\cite{1d}, the solution of  eq.(\ref{SDDF}) is given by  $\rho(z')=\rho_0+A \exp(- \bar c z')$ at the  {\it front}, where $A/\rho_0$ is a constant, fixed by the nature of the  barrier and the boundary conditions, and by   $\rho(z')=\rho_0$, without any kind of structure, in the  {\it wake}  region. It may be proved by a formal  reduction of the stationary DDF equation  with $\Delta {\cal F}=0$  in any dimension, that this exponential decay behaviour at the {\it front}, and this absence of excess molecules in the {\it wake}, are generic characteristics of the  total excess density distribution, under constant bulk boundary conditions ($\lim_{ |{\bf r'}|\rightarrow\infty}\rho= \rho_{o}$), for the integrated  excess  density distribution across the transverse directions, i.e.  $\frac{\Delta n(z')}{\rho_{o}}=\frac{1}{\sigma}\int{dx (\frac{\rho({\bf r'})}{\rho_{o}}-1)}$  in effective $2D$  systems, and $\frac{\Delta N(z')}{\rho_{o}}=\frac{2\pi}{\sigma^{2}}\ \int{Dr R (\frac{\rho({\bf r'})}{\rho_{o}}}-1)$  in effective $3D$ systems.\\ 
 In effective $2D$ systems, for a shifting  cylindric   external potential (e.g. the dielectric potential  created by a laser beam through a bulk solution), eq.(\ref {SDDF}), for the region with  $V_{{\rm ext}}({\bf r'})=0$,  reduces  to,
\begin{equation}
\frac{\partial^{2}\rho }{\partial x ^{2} }+\frac{\partial^{2}\rho}{\partial z'^{2}}+\bar{c}\, \frac{\partial\rho}{\partial z'}=0.
\end{equation}
Through a Fourier transform, the solution is given by,
\begin{equation}
\rho (x ,z')= \rho_{0}+ \int_{0}^{\infty} d\alpha \, \cos[\alpha x ]\, e^{-\beta z'} g(\alpha);
\end{equation}
where,
\begin{equation}
\beta_{\pm}=\frac{\bar{c}}{2}\pm\sqrt{\left(\frac{\bar{c}}{2}\right)^{2}+\alpha^{2}}.
\end{equation}
\begin{figure}
\vspace{-2.5cm}
\includegraphics[width=130mm]{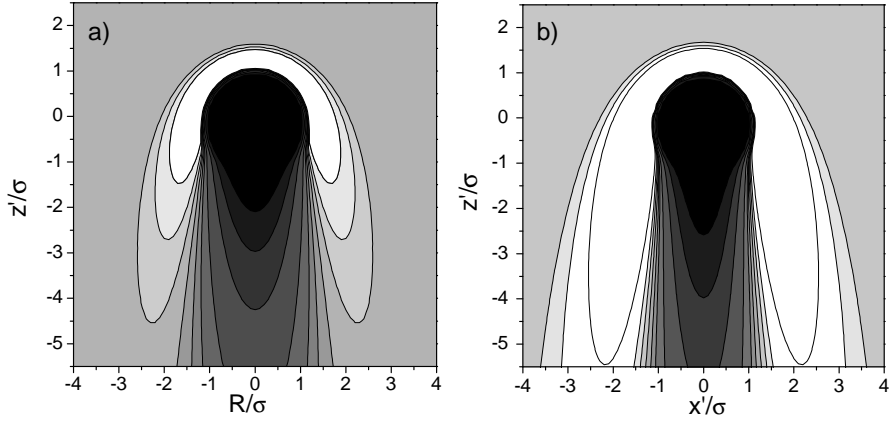}
\vspace{-0.5cm}

\caption[]{ Steady state contour density field of non interacting colloidal  particles a$)$ for an  effective   $3D$ system  and  b$)$ for an  effective   $2D$ system, dragging at constant velocity $\sigma \bar{c}=10$ by an external potential  $V_{ext}=V_{o} e^{-|{\bf r'}/\sigma|^{6}}$ with $\beta V_{o}=10$. Dark regions correspond to low densities  while the  bright one show high densities.}
\label{fig:1}
\end{figure}

Far away from the external potential, a fast convergence of the Fourier 
components $g(\alpha)$  is observed, and  the relevant features come from
their behavior for \begin{math}\alpha\ll\bar{c}\end{math}. Hence we may
use  \begin{math}
\beta_{+}\approx\bar{ c }+\frac{\alpha^{2}}{\bar{ c }}\end{math} and 
\begin{math} \beta_{-}\approx -\frac{\alpha^{2}}{\bar{ c }}\end{math}, 
for the  {\it front}  and   {\it wake}  regions respectively. The expansion of 
$g_{\pm}(\alpha)$ 
as an even  polynomial function for small $\alpha$, and the zero {\it wake}  
requirement, lead in the  {\em wake} to 
$ g_{-}(\alpha)\approx A_{1}\alpha ^{2}+ A_{2} \alpha ^{4}+...$, while at the {\it front}  we expect  
$g_{+}(\alpha)\approx B_{0}+ B_{1}\alpha ^{2}+ B_{2} \alpha ^{4}+... $
Thus, 
\begin{eqnarray}
\rho(x,z')\approx \rho_{0} +\sqrt{\pi} e^{-\bar{c} z'}e^{-\frac{a^{2}}{4}}\left[B_{0}\frac{w}{2}
+B_{1}\frac{ w^{3}}{8}\left(a^{2}-2\right)
+ B_{2}\frac{ w^{5}}{32} \left(a^{4}-12 a^{2}+12\right)+... \right],
\label{2Dsol1}
\end{eqnarray}
for $z'\gg \sigma $, and
\begin{eqnarray}
\rho(x,z')\approx \rho_{0} + \sqrt{\pi} 
e^{-\frac{a^{2}}{4}} \left[A_{1}\frac{w^{3}}{8}\left(a^{2}-2\right )
+ A_{2}\frac{ w^{5}}{32} \left(a^{4}-12a^{2}+12\right)+... \right];
\label{2Dsol3}
\end{eqnarray}
for $z'\ll -\sigma $;
where
\begin{math}
w=\sqrt{\bar{c}/\mid z' \mid} \,\,\end{math} 
and\begin{math}\ \, \,\,a=wx \end{math}.\\ 

\begin{figure}
\includegraphics[width=130mm]{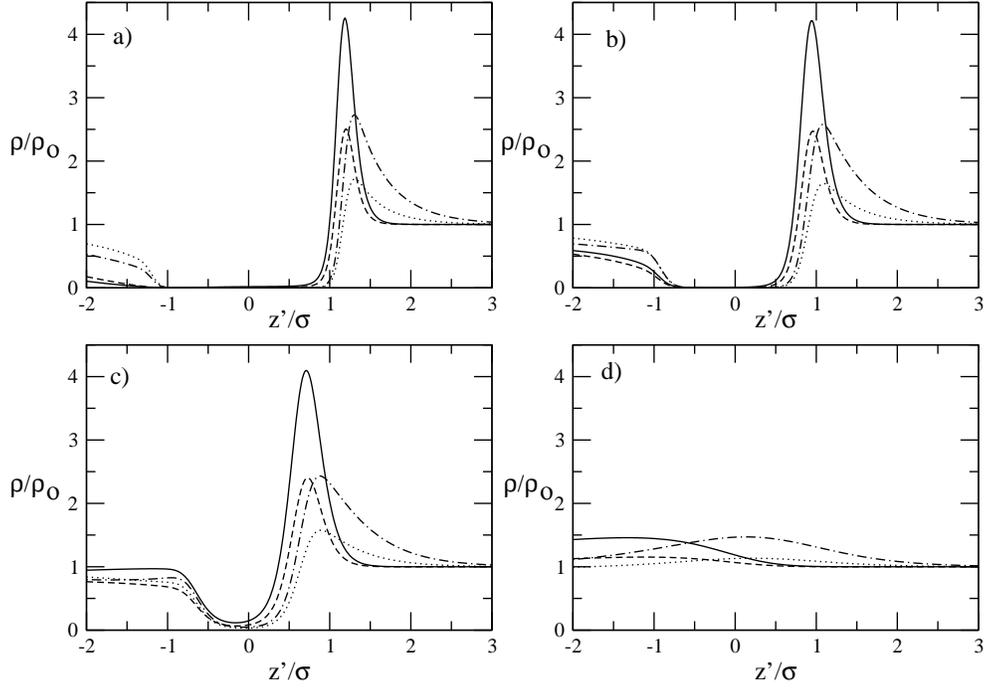}
\caption[]{Steady state density profile of colloidal particles  under the influence of an external potential  $V_{ext}=V_{o} e^{-|{\bf r'}/\sigma|^{6}}$, with $\beta V_{o}=10$, for  effective  $2D$ system at drift velocity  $\sigma\bar{c}=10$ (solid line)  and $\sigma\bar{c}=2$ (dotted-dashed line) and for effective  $3D$ system at drift velocity  $\sigma\bar{c}=10$ (dashed line) and $\sigma\bar{c}=2$ (dotted line). The fixed radial  and perpendicular distance to the z'axis are a$)$ $R=x=0$, b$)$ $R=x=0.75\sigma$, c$)$ $R=x=\sigma$, d$)$ $R=x=2\sigma$.  }
\label{fig:2}
\end{figure}

In effective  3D system, for a shifting spherical external potential (e.g. a driven colloid particle perturbing a bulk solution of others colloids), we use cylindrical coordinates ${\bf r}'=(R,\phi,z')$, and eq.(\ref{SDDF}), for the region with  $V_{{\rm ext}}({\bf r'})=0$, reduces  to,

\begin{eqnarray}
\frac{1}{R}\frac{\partial}{\partial R }\left(R \frac{\partial \rho({\bf r'})
}{\partial R }\right)+
\frac{\partial^{2} \rho({\bf r'})
}{\partial z'^{2} }+\bar{c}\frac{\partial \rho({\bf r'})
}{\partial z'}=0.
\label{3Deq}
\end{eqnarray}
Through a  Hankel transform, the solution  is given by ,
\begin{eqnarray}
\rho(R,z')=\rho_{0}+\int_{0}^{\infty}d\alpha \,\alpha\,f(\alpha)\,
  J_{0}[\alpha\, R] \, e^{-\beta \, z'};
 \label{3Dsol}
\end{eqnarray}
where $ J_{0}$ is the zeroth order Bessel function. Proceeding with the Hankel components as in the effective $2D$ 
with the Fourier ones, we might take  $ f_{-}(\alpha)\approx C_{1}\alpha ^{2}+ C_{2} \alpha ^{4}+...$ 
in the  {\it wake}, and  at the  {\it front} $f_{+}(\alpha)\approx D_{0}+ D_{1}\alpha ^{2}+ D_{2} \alpha ^{4}+... $.  
Thus, 
\begin{eqnarray}
\rho(R,z')\approx \rho_{0} + e^{-\bar{c} z'}e^{-\frac{h^{2}}{4}}\left[D_{0}\frac{w^{2}}{2}
+D_{1}\frac{ w^{4}}{8}\left(h^{2}-4\right)
+ D_{2}\frac{ w^{6}}{32} \left(h^{4}-16 h^{2}+32\right)+... \right],
\label{3Dsol1}
\end{eqnarray}
for $z'\gg \sigma $, and
\begin{eqnarray}
\rho(R,z')\approx \rho_{0} + 
e^{-\frac{h^{2}}{4}} \left[C_{1}\frac{ w^{4}}{8}\left(h^{2}-4\right )
+ C_{2}\frac{ w^{6}}{32} \left(h^{4}-16 h^{2}+32\right)+... \right];
\label{3Dsol3}
\end{eqnarray}
for $z'\ll -\sigma$;
where
\begin{math}
h=wR\end{math}.\\ 
\begin{figure}
\includegraphics[width=130mm]{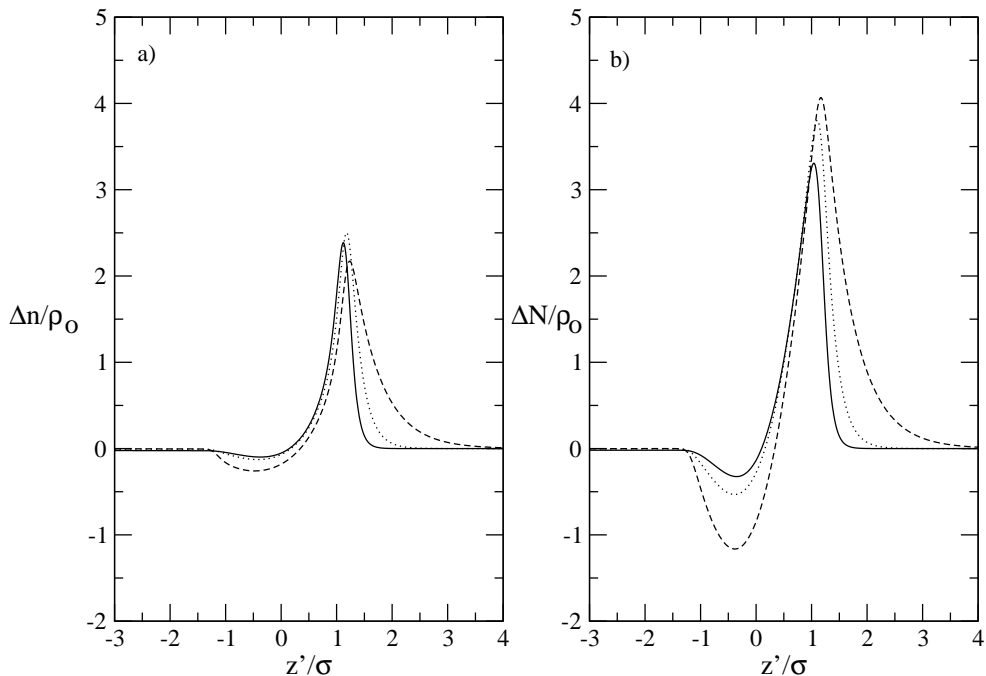}
\caption[]{Integrated excess  of the stationary density distribution over the transverse directions, a$)$ for an effective  $2D$   and b$)$ $3D$ system. The  external potential $V_{ext}=V_{o} e^{-|{\bf r'}/\sigma|^{6}}$, with $\beta V_{o}=10$ is shifting at the  velocities  $\sigma\bar{c}=10$ ( solid  lines),$\sigma\bar{c}=5$ ( dotted  lines) and  $\sigma\bar{c}=2$ ( dashed  lines).}
\label{fig:3}
\end{figure}

The coefficients (A,B,C and D) are determined by the particular choice of the external potential but, for each dimension, the asymptotic decay forms are generic.  In the transverse plane (for  fixed  $z'$) the excess stationary density distribution ( $\rho({\bf r'})-\rho_{o}$) is  given by a gaussian times a polynomial function, and  along the direction of the  movement (z' axis) for a fixed distance to the origin ($a=const$ in effective $2D$ systems  and $h=const$ in  the $3D$ ) is  given, at the  {\it front}, by an exponential decay ($ \exp(-\bar{c} z')$)   and an inverse power in the  {\it wake}.
Over the transverse directions, at the  {\it front},  the maximum density  is at the  $z'$ axis, while in the  {\it wake},  this axis is a minimum  between two symmetric maxima which produce a  cap-like structure with  parabolic shape. The distance between the  maximum and the z' axis is  $  2 \, \sqrt{2 \mid z' \mid/\bar{c}}\sigma$  in  effective $3D$ systems and  $  \sqrt{6 \mid z' \mid/\bar{c}}\sigma$ in the $2D$. In fig.1 we present the numerical solution for the stationary density distribution of colloidal particles under the influence of  an external potential $V_{ext}=V_{o} e^{-|{\bf r'}/\sigma|^{6}}$, with $\beta V_{o} =10$, shifting at the velocity $\bar{c}\sigma=10$, in  a$)$  an effective $3D$  and b$)$  $2D$ system. The clear cap-like structure, extended further away from the external potential, is  clearly observed. 

We have mentioned before  that only  in effective $1D$ systems the {\it front}  and  {\it wake} are fully separated by the external potential,  which implicate that as the solution of eq.(\ref{SDDF}),  the {\it front}  and  {\it wake} should share   an unique analytic expression   in effective  dimensions beyond $1D$.
This unique expression produces, at the  {\it front} ({\it wake}),  an exponential growth  of the $ \beta_{+}(\beta_{-})$ contribution, which only by the appropriate behaviour of $g(\alpha)(f(\alpha))$ for large $\alpha$ be canceled. We have compared our analytical prediction with the numerical solution of eq.(\ref{SDDF}), the concordance reveal a local asymptotic convergence which makes useful those asymptotic expressions (eqs.(\ref{2Dsol1}),(\ref{2Dsol3}),(\ref{3Dsol1}),(\ref{3Dsol3})), to represent separately  (at the  {\it front} and in the {\it wake} regions)  the stationary state. The use 
of polar (spherical) coordinates, instead of the rectangular (cylindric) ones, would not solve the problem, because the parabolic structure of the {\it wake} implies the entanglement of the radial and the angular coordinates.\\
Because this work may be usefully in the construction of particles pumps, we want to point out some effects to be considered in the  design of an efficient device  to drag particles. For the same radial form of the  external potential 
it is clear that an effective  $2D$ system is more efficient dragging particles than a $3D$,  because it produces a  higher density at the  {\it front}  peak. In fig.2 we present the numerical stationary density distribution of an effective $3D$  (dashed and dotted  lines) and $2D$ system (solid  and dotted-dashed lines), for different  fixed values of x and R. The higher {\it front} density peak in the effective $2D$ system is clearly observed. For the same reason, an effective $1D$ system is more efficient than the $2D$. 

For an effective  $1D$ system we obtained in our previous  work ref.\cite{1d}  a non-linear dependence of the total current ($j=\bar{c} \int{(\rho({\bf r'})-\rho_{o})d{\bf r'} }$) with the drift velocity. If we denoted by 
$\bar{c}_{max}$, the velocity at which the total current is maximum   for a given external potential, the efficiency of a device as a particles pump increases with  $\bar{c}$   until   $\bar{c}_{max}$  is reached, and then diminishes for higher   $\bar{c}$.
As we have mentioned before,  the integrated  excess of the stationary density distribution across the transverse directions, $\Delta n $  in effective $2D$ systems and $\Delta N $  in the  $3D$, satisfies the same stationary DDF equation as the $1D$  density distribution, so we may expect for  the total current  this non-linear behaviour too. To support numerically our analytical result  in fig.3 we present  the  integrated  excess of the stationary density distribution,  a$)$   in an effective   $2D$ and  b$)$  $3D$ system, the exponential decay behaviour at the {\it front} and the absence of structure in the {\it wake} as for the $1D$ stationary density distribution is clearly observed. We have estimated that the maximum total currents are obtain for  $\sigma \bar{c}_{max} $   $\approx 0.5$  in $1D$, $\approx 12.41$  in $2D$, and   $\approx 14.76$  in $3D$, which implicate that, apart from the dependence of  $\bar{c}_{max} $ with the external potential shape, it is a function of the effective system dimension. $\bar{c}_{max} $ determines when for a given external potential the system is saturated by the drift velocity. Increasing the dimension, we increase the paths for the molecules to avoid the external potential so a higher velocity is needed to make the molecules go thought the external potential.\\

  We have considered the geometrical effects of the effective dimension
in systems of non-interacting colloidal particles driven by external barriers
with constant drift rates.  The main difference between the 2D and 3D cases
analyzed here and the previously reported 1D case is the appearance of
long wake structures behind the moving barrier. In contrast with the
common exponential decay $\rho({\bf r'})-\rho_o \sim exp(-\bar{c} z')$ for the
structure at barrier {\it front}, the structure of the  {\it wake} decays only
as an inverse power, $\rho({\bf r'})-\rho_o \sim 1/|z'|^2$ for 3D and 
$\rho({\bf r'})-\rho_o \sim 1/|z'|^{3/2}$ for 2D. However, the transverse integral
of the wake, over the directions other than $z'$, gives no excess
of particles, with respect to the unperturbed density $\rho_o$ far from
the potential barrier. This is in agreement with the complete lack 
of {\it wake} structure behind a 1D barrier, and it implies that the 
global current induced by the shifting potentials, calculated as the
integral of $\bar{c} (\rho({\bf r'})-\rho_o)$ comes from the balance between
the positive contribution from the density peak at the {\it front}, and
the void ($\rho({\bf r'}) < \rho_o$) within the barrier. This total current 
follows a non-monotonic dependence with the shifting rates,
as reported in 1D \cite{1d}, with an optimum shifting rate $\bar{c}_{max}$ which
increases with the effective dimensions. The colloid-colloid
interactions effects, which have being leaved out of this work,
may be included within the general framework of the DDF \cite{1d,joe2,prlfrus}
and may help to the design of efficient particle pump devices.\\

F. P  gratefully acknowledges discussions with J.D.
This work has been supported by the Direcci\'on General de
Investigaci\'on Cient\'{\i}fica (MCyT) under grant BMF2001-1679-C03-02, 
and a FPU grant AP2001-0074 from the MECD of Spain. 



\end{document}